\documentclass[aps,prb,10pt,twocolumn,amsmath,amssymb,superscriptaddress]{revtex4-2}

\usepackage{graphicx}
\usepackage{amsmath}
\usepackage{dcolumn}
\usepackage{bm}
\usepackage{xcolor}
\usepackage{graphicx}
\usepackage{physics}
\usepackage{mathptmx}
\usepackage{lipsum}
\usepackage{subfigure}
\usepackage{fixltx2e}
\usepackage[version=4]{mhchem} 
\usepackage[cal = cm]{mathalpha} 
\begin{document}

\title{Magnetic order in the computational 2D materials database (C2DB) from high throughput spin spiral calculations}

\author{Joachim Sødequist}
 \affiliation{Computational Atomic-Scale Materials Design (CAMD), Department of Physics, Technical University of Denmark, 2800 Kgs. Lyngby, Denmark}
\author{Thomas Olsen}
\email{tolsen@fysik.dtu.dk}
\affiliation{Computational Atomic-Scale Materials Design (CAMD), Department of Physics, Technical University of Denmark, 2800 Kgs. Lyngby, Denmark}

\date{\today}

\begin{abstract}
We report a detailed investigation of the magnetic order in 192 stable magnetic two-dimensional materials from the Computational 2D Materials Database having one magnetic atom in the unit cell. The calculations are based on a systematic workflow that employs spin spiral calculations and yields the magnetic order in terms of a two-dimensional ordering vector $\mathbf{Q}$. We then include spin-orbit coupling to extract the easy and hard axes for collinear structures and the orientation of spiral planes in non-collinear structures. Finally, for all predicted ferromagnets we compute the Dzyaloshinskii-Moriya interactions and determine whether or not these are strong enough to overcome the magnetic anisotropy and stabilise a chiral spin spiral ground state. These steps completely determines the ground state order within the spiralling ansatz. We find 58 ferromagnets, 21 collinear anti-ferromagnets, and 85 non-collinear ground states of which 15 are chiral spin spirals driven by Dzyaloshinskii-Moriya interactions. The results show that non-collinear order is in fact as common as collinear order in these materials and emphasise the need for detailed investigation of the magnetic ground state when reporting magnetic properties of new materials. Furthermore, non-collinear order typically breaks symmetries inherent to the lattice and may give rise to emergent properties such as multiferroicity, magnetoelectricity or second order optical effects that would be predicted as absent based on a collinear assumption.
\end{abstract}

\maketitle

\section{Introduction}\label{sec:intro}
The discovery of ferromagnetic order in monolayer CrI$_3$ \cite{Huang2017} has initiated a huge interest in magnetism of two-dimensional (2D) materials \cite{Burch2018, Huang2018b,Liu2018f,Miyazato2018,Gong2019,Olsen2019,Sethulakshmi2019,Li2019,Torelli2019b,Torelli2020,Torelli2020a,Shen2022}. The most intriguing aspect of the field is arguably the fact that the fundamental mechanism that drives magnetic ordering in 2D is fundamentally different from the case of three dimensions (3D), where order arises from spontaneous symmetry breaking. In 2D this is not allowed because the entropic gain in free energy always favours disorder at finite temperatures \cite{Mermin1966}. As such, magnetic order in 2D relies on spin-orbit coupling and magnetic anisotropy to break the continuous spin rotational symmetry explicitly \cite{Lado2017,Torelli2018,Lu2019}. For the case of CrI$_3$, the magnetic order is driven by a strong out-of-plane easy axis \cite{Huang2017} whereas measurements on CrGeTe$_3$ has shown that magnetic order persists for exfoliated bilayers, but vanishes in the monolayer limit due to an easy-plane \cite{Gong2017a}, which retains part of the continuous spin rotational symmetry. Since the seminal works on CrI$_3$ and CrGeTe$_3$, several other 2D materials has been shown to exhibit magnetic order. CrBr$_3$ was demonstrated to order below 21 K as a result of easy-axis anisotropy \cite{Jin2020}, whereas CrCl$_3$ exhibits Kosterlitz-Thouless physics, but no strict long range order \cite{Bramwell1993} due to easy-plane anisotropy that does not fully break spin rotational symmetry \cite{Bedoya-Pinto2021}. In addition, anti-ferromagnetic order has been studied in monolayers of the transition metal phosphorous sulphides (APS$_3$, A=Co, Fe, Ni, Mn), which comprises a versatile class of materials containing both Néel and stripy ordered anti-ferromagnets as well as easy-plane and easy-axis anisotropy \cite{Jernberg1984,Joy1992,Wildes1998,Wildes2015,Du2016,Lee2016,Wang2016h,Wang2018,Kim2019,Olsen2021,Kim2021}. Finally, Fe$_3$GeTe$_2$ comprises an example of a metallic ferromagnetic monolayer, which has a strong easy-axis and orders up to 120 K \cite{Fei2018}. Subsequently, monolayers Fe$_4$GeTe$_2$ \cite{Seo2020} and Fe$_5$GeTe$_2$ \cite{Chen2022} have both been shown to exhibit Curie temperatures close to room temperature. Metallic compounds thus appear to be a promising venue for realising high intrinsic ordering temperatures in 2D - possibly due to long range exchange interactions. But more importantly, the magnetic properties may be tuned by applying an external gate voltage that controls the position of the Fermi levels and metallic ferromagnets thus comprises a highly versatile platform for manipulating spins in 2D.
\begin{figure*}[tb]
    \centering
    \includegraphics[width=\linewidth]{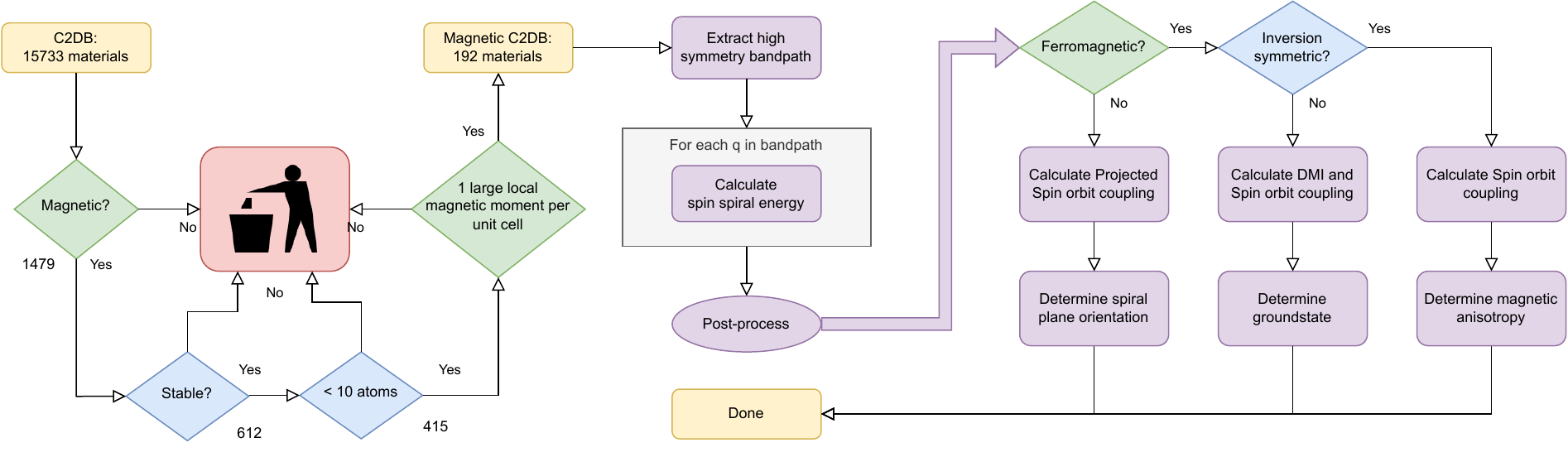}
    \caption{The workflow for selecting and processing magnetic properties of materials in the C2DB.}
    \label{fig:workflow}
\end{figure*}

In general, ferromagnetism, and anti-ferromagnetism constitutes limiting cases of non-collinear spiral magnetic order \cite{Kaplan1959}. The magnetic order is then described by an ordering vector $\mathbf{Q}$, which encodes how the magnetisation rotates under lattice translations. Collinear order corresponds to $\mathbf{Q}$ being situated at the Brillouin zone centre or at certain high symmetry points, but non-collinear order with general $\mathbf{Q}$ (which may even be incommensurate with the lattice) are commonly found in 3D magnets. Recently, monolayers of NiI$_2$ were found to exhibit such incommensurate spiral order \cite{Song2022} and first principles calculations have subsequently predicted that spiral order appears commonly in isostructural transition metal halides \cite{Sødequist2023}. The presence of spiral order have large consequences for the physical properties of a given material, since it typically breaks part of the lattice symmetry. In particular, for the case of NiI$_2$ and related transition metal halides, the spiral order introduces a polar axis, which gives rise to ferroelectric order. Such type II multiferroics imply that the magnetic order may be controlled by external electric field \cite{Matsukura2015}.

Determining the ground state magnetic order of a material can be a challenging task. Experimentally, the magnetic structure is typically obtained from neutron diffraction experiments, where the observed Bragg peaks are compared with the expectation from different ordered structures and the best fit is identified as the most likely ground state structure \cite{Ressouche2014}. For 2D materials, it is not possible to perform neutron scattering experiments and one has to rely on Raman spectroscopy \cite{Wang2016h,He2018} or second harmonic generation \cite{Sun2019}, from which it is much more difficult to extract the detailed magnetic structure. From a theoretical point of view, density functional theory (DFT) will typically provide accurate predictions on the magnetic structure \cite{Illas1998,Xiang2013,Olsen2017}, but standard super cell approaches are limited by the fact that one has to compare total energies of all possible magnetic configurations. Even for collinear anti-ferromagnets, this can be an insurmountable task if there is many configurations and it is in general impossible to predict incommensurate spin spiral order from a super cell approach. These limitations may be remedied by application of the generalized Bloch theorem (GBT) \cite{Sandratskii1986,Heide2009,Lezaic2013,Sødequist2023}, where the spiral order is explicitly included in rotating boundary conditions for the magnetisation upon lattice translations. This implies that any spiral order can be simulated in the primitive unit cell, which greatly simplifies the determination of the ground state magnetic order.

In the present work we have set out to determine the ground state magnetic order of all stable magnetic 2D materials in the Computational 2D Materials Database (C2DB) \cite{Haastrup2018,Gjerding2021} containing one magnetic atom. The C2DB currently contains 15733 2D materials, which includes experimentally known compounds as well as 2D materials obtained from theoretical exfoliation of 3D bulk materials \cite{Torelli2020a} and compounds obtained from systematic combinatorial lattice decoration of common prototypes \cite{lyngby2022data}. All materials in the C2DB are generated with ferromagnetic order, but the true magnetic order is not known. In fact,  explicit calculations of the nearest neighbour exchange interaction have directly shown that the ferromagnetic configuration cannot be the ground state in many of the materials \cite{Torelli2019b,Torelli2020a}. Here we apply the GBT to extract the true magnetic order (as predicted by DFT). The magnetic order is then characterised by two parameters: an ordering vector $\mathbf{Q}$ and a unit vector $\mathbf{\hat n}$ that provides either the easy-axis (for ferromagnets) or the normal vector to the spiral plane (for spiral magnets). The preferred direction of spin is a spin-orbit effect, which we include as a post-processing step. Finally, for all predicted ferromagnets lacking inversion symmetry, the true ground state may be a long wavelength chiral spin spiral driven by Dzyaloshinskii-Moriya (DM) interactions \cite{Heide2009}. We thus calculate the DM interaction for all these materials and derive a criterion for whether or not the DM energy is able to overcome the magnetic anisotropy energy that is lost in the chiral spin spiral state. All results have been added to the C2DB and can be accessed freely from the online web interface.

The paper is organised as follows. In Sec. \ref{sec:workflow} we describe the details of our workflow and the computational details that were used to predict the magnetic ground states. In Sec. \ref{sec:results} we first exemplify the procedure used to obtain ordering vectors, spin orientation and DM interaction for specific materials. We then summarise the statistics of out results with respect to occurrence of various types of magnetic order and finally comment on the magnon dispersions, which may be extracted from spin spiral calculations. In Sec. \ref{sec:discussion} we provide a discussion and outlook

\section{Computational details and workflow}\label{sec:workflow}
The DFT calculations were performed using the open source electronic structure code GPAW \cite{Enkovaara2010}, using the projector augmented wave method \cite{blochl} and a plane waves with a cutoff of 800 eV. All calculations was carried out in the GBT approach using 
the local density approximation (LDA) and a gamma-centred $k$-point grid with a density of 6 {\AA}. For each material we performed GBT calculations for $q$-points along a standardised path in the Brillouin zone (specified by the 2D Bravais lattice) \cite{Setyawan2010} sampling 10 equidistant points along each line segments of the path. The atomic structures and magnitude of initial magnetic moments were taken from the C2DB (optimised ferromagnetic structure with the Perdew-Burke-Ernzerhof (PBE) xc-functional \cite{pbe}) and the  magnetic moments were then rotated to fit the spin spiral boundary conditions $\mathbf{m} = m [\sin{\mathbf{q}\cdot\mathbf{a}}, \cos{\mathbf{q}\cdot\mathbf{a}}, 0]$. The workflow is implemented in the Python framework Atomic Simulations Recipes (ASR) \cite{gjerding2021atomic}, which allow for a flexible and modular workflow, while ensuring dependencies are properly handled and documented. Thus, we are able to treat branching of the workflow as the partial results on the magnetic ground state becomes available. We combine the ASR workflow together with the Python front-end for MyQueue \cite{Mortensen2020}, which further allows for state-dependent job scheduling. The ASR package is interfaced to the Atomic Simulations Environment (ASE) \cite{Larsen2017} in order to handle database level features such as selecting screening criteria, gathering results and merging the resulting databases. 

The workflow consists of four main steps and is illustrated in figure \ref{fig:workflow}. The first step is pre-screening of C2DB, where we select materials that were found to have a stable ferromagnetic structure with PBE. The stability requirement involves both dynamical stability criterion (no imaginary gamma-point phonon frequencies in a $2\times2$ super cell) and a thermodynamical stability criterion where we only include materials with a formation energy within 0.2 eV/atom of the convex hull (the convex hull contains structures from C2DB and bulk 3D materials from OQMD \cite{Saal2013}). 

In the second step, the workflow calculates the flat spin spiral energy for $\mathbf{q}$-points along a 2D Brillouin zone high-symmetry path and the ground state ordering vector $\mathbf{Q}$ is then found by a parabolic fit in the vicinity of the $q$-points with minimum energy. The energy as a function of $\mathbf{q}$ is in general expected to be smooth. However, due to the added boundary constraints of the GBT, calculations far from the ground state ordering vector may lead to convergence problems or produce a non-magnetic state with much larger energy. The workflow proceeds regardless as long as the ground state is properly converged and magnetic.


In the third step we determine the orientation of the spin spiral normal vector $\hat{\mathbf{n}}$ with respect to the lattice and the ordering vector $\mathbf{Q}$. This is done by including spin-orbit coupling (SOC) non-selfconsistently and calculating the energy for all non-equivalent orientations (represented by equidistant points on the upper hemisphere). For ferromagnets ($\mathbf{Q}=(0,0)$) we obtain the easy axis as the direction $\mathbf{\hat n}$ that minimizes the energy. For spin spiral ground states, the spin-orbit interaction is not compatible with the GBT and we apply the projected spin-orbit interaction, where only the components of angular momentum orthogonal to the spin plane is included. Such a term yields the only contribution in first order perturbation theory \cite{Heide2009} and has previously been shown to compare well with super cell simulations of spin spirals including full spin-orbit coupling \cite{Sandratskii2017, Sødequist2023}. For spiral ordered compounds we thus report $\mathbf{\hat n}$ as the normal vector to the spin spiral plane that minimises the energy. The ordering vectors will be stated in units of reciprocal lattice vectors and the orientation $\mathbf{\hat n}=(\theta,\varphi)$ is given in terms of the polar angle $\theta$ relative to the out-of-plane direction and an azimuthal angle $\varphi$ that measures the deviation from the $x$-axis. 

The final step of the workflow checks if a given material is a ferromagnet lacking inversion symmetry. If that is the case the energy may be lowered by forming a chiral long wavelength spin spiral driven by DM interactions. We then calculate the DM interaction in the vicinity of $\mathbf{Q}=(0,0)$ and determine if the energy gain of a chiral spin spiral is sufficient to overcome a possible loss of magnetic anisotropy. As a result, some of the materials classified as ferromagnets in step two become classified as chiral spin spirals. It should be noted that the DM interaction may also have an effect on materials that are predicted to have a spiral ground state without spin-orbit coupling, but in such cases the DM interaction typically only gives rise to small modifications of the ordering vector $\mathbf{Q}$ (which is already finite) and we have not included these effects in the workflow.


\begin{figure*}
    \centering
    \includegraphics{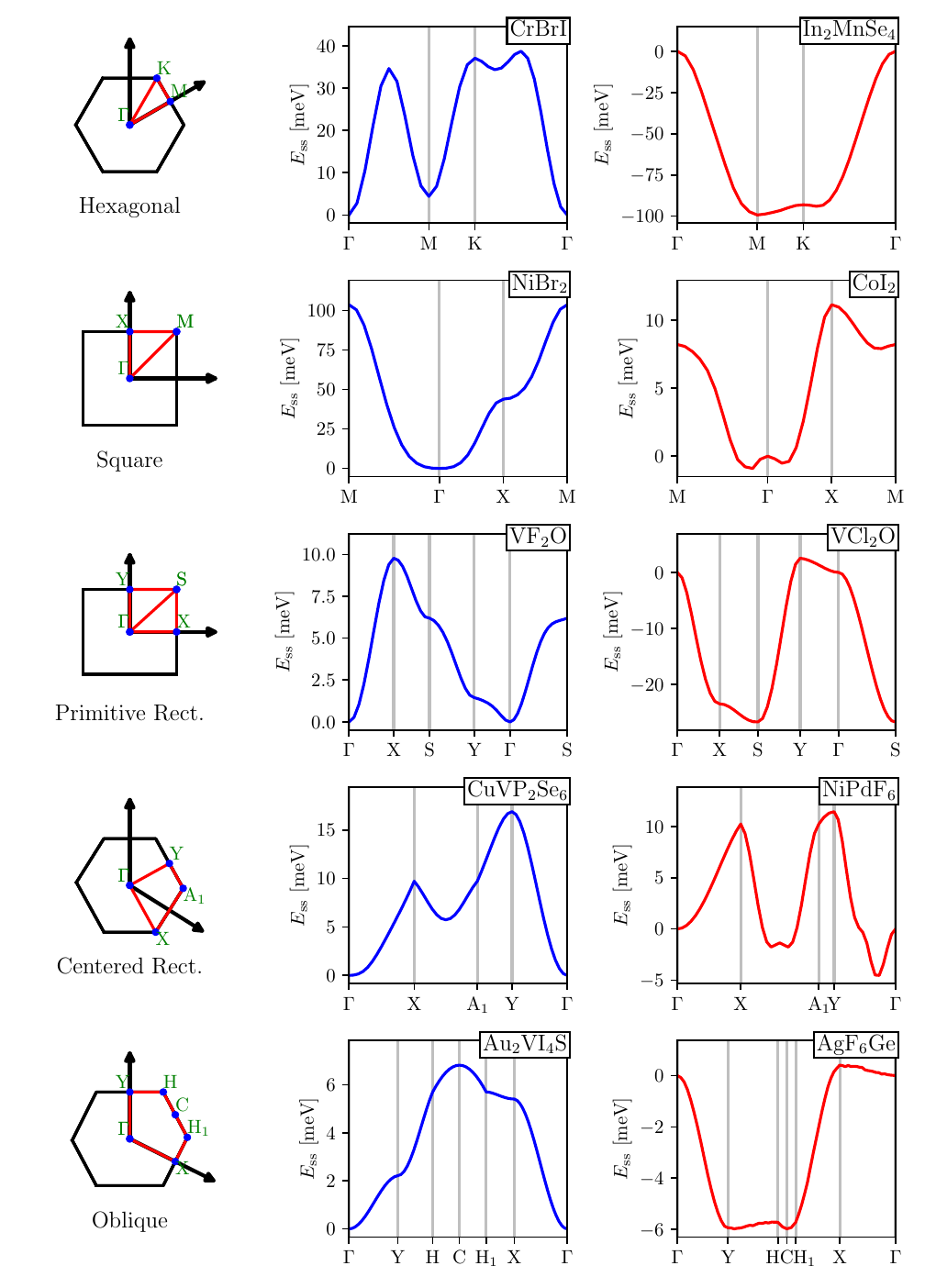}
    \caption{Examples of spin spiral dispersions of hexagonal, square, primitive rectangular, centered rectangular and oblique Bravais lattices. The spiral dispersions to the left (shown in blue) have minima in $\Gamma$ and imply ferromagnetic ground states. The spiral dispersions to the right (shown in red) do not have minima in $\Gamma$ implying non-ferromagnetic ground states that may be collinear antiferromagnets (minima in X, Y, M, S, or C) or non-collinear spin spirals (minima at at other points in the BZ). }
    \label{fig:examples}
\end{figure*}

\section{Results}\label{sec:results}
\subsection{Magnetic order from spiral calculations}
We start by exemplifying the various types of magnetic orders that are obtained from the spin spiral calculations and in figure \ref{fig:examples} we show examples of spin spiral dispersions in the five 2D Bravais lattices. 
The magnetic ground states can be categorised into ferromagnetic, collinear antiferromagnetic and non-collinear spin spirals.
Ferromagnetic compounds are grouped in the middle column
where the minimum energy is located at $\mathbf{Q} = \Gamma$ in each Bravais lattice. In the right column we show collinear antiferromagnetic order in In$_2$MnSe$_4$, VCl$_2$O and AgF$_6$Ge where the order parameters are $\mathbf{Q} = (1/2, 0)$, $(1/2, 1/2)$ and $(1/2, -1/2)$ respectively. In addition we show non-collinear spin spirals in CoI$_2$ with $\mathbf{q} \simeq (0.08, 0.08)$ and NiPdF$_6$ with $\mathbf{q} = (0.13, -0.13)$. 
In addition to the magnetic ground state, the spin spiral dispersion may provide additional information, such as the spin stiffness, which is closely related to the curvature at the minimum. In particular, the case of NiBr$_2$ exhibits a rather low curvature at $\Gamma$, which makes it particularly prone to destabilise into a spin spiral ground state due to DM interactions. We return to the quantitative analysis of this issue in section \ref{sec:dmi}. Another point of interest is the presence of additional local energy minima that comes close to the global minimum and may imply proximity to a magnetic phase transition. For example, the case of CrBrI shown in figure \ref{fig:examples} has a local minimum at $\mathbf{q}=\mathrm{M}$ which is located roughly 5 meV above the global minimum. Such a local minimum could transition to a global minimum under external perturbations (e.g. strain) or the magnetic order may change at elevated temperatures due to larger entropic contributions at the local minimum. In addition, the local exchange interactions is commonly analysed in terms of different spin configurations based on super cell calculations, but such results may be misleading if the high symmetry states are local minima. 
Finally, in isotropic materials exhibiting spin spiral ground states, we often find one-dimensional near-degenerate manifolds. In figure \ref{fig:examples} this is exemplified by quadratic CoI$_2$, which is characterised by a near degenerate ring encircling $\Gamma$ - reminiscent of a spiral spin liquid at finite temperature \cite{gao2022spiral}. In strongly anisotropic materials, the exchange interactions may be rather weak along particular lattice vectors, which yields vanishing dispersion along the weakly coupled direction. This can be seen in figure \ref{fig:examples} in oblique AgF$_6$Ge, where the line between H and Y is nearly degenerate. 
\begin{figure*}
    \centering
    \includegraphics{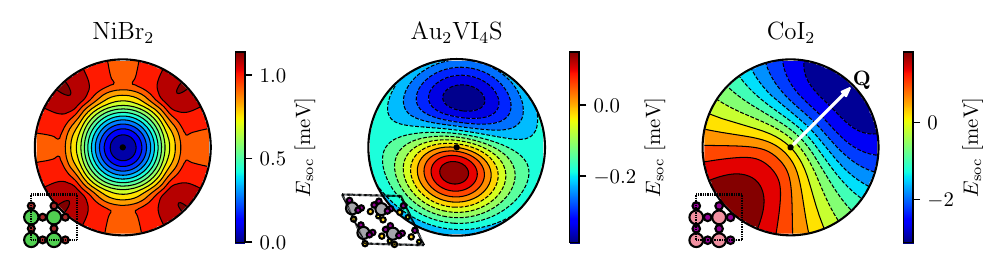}
    \caption{Magnetic anisotropy energies relative to the out-of-plane direction of two ferromagnets (Au$_2$VI$_4$S and NiBr$_2$) and the spin spiral CoI$_2$ where the direction of the ordering vector is indicated by the white arrow. The radial axis corresponds to $\theta$ where the center black dot denotes the out-of-plane direction and the outer perimeter the in-plane direction. For the ferromagnets the direction indicate the spin alignment whereas the direction of the spin spiral is the normal vector to the spiral plane. The insets show 2x2 super cells of the respective structures.}
    \label{fig:soc_examples}
\end{figure*}

\subsection{Magnetic anisotropy}
The orientation of the spins relative to the lattice can be determined by minimising the spin-orbit energy. 
In figure \ref{fig:soc_examples} we show of the spin-orbit energies (relative to the $z$-direction) in NiBr$_2$, Au$_2$VI$_4$ and CoI$_2$ represented as stereographic projections on the atomic plane. The case of NiBr$_2$ is an easy-axis (out-of-plane) ferromagnet with quadratic symmetry. In the vicinity of the easy-axis the structure is isotropic as expected from the four-fold rotational symmetry. However, the isotropy is weakly broken for spins aligned with the atomic plane and gives rise to two equivalent hard axes. This effect cannot be captured in a simple Heisenberg description with quadratic spin interactions and signals the presence of higher order anisotropic interactions. We return to this point in section \ref{sec:dmi}. In contrast, the ferromagnet Au$_2$VI$_4$S is devoid of any symmetries and has distinct hard and easy axes.
Finally, CoI$_2$ exhibits spiral order with ordering vector $\mathbf{Q} = (0.08, 0.08)$ and we find that the spin-orbit energy is minimised when the normal vector of the spiral plane $\mathbf{\hat{n}}$ is parallel with $\mathbf{Q}$, corresponding to a proper screw.

\subsection{Chiral spin spirals from DM interactions}\label{sec:dmi}
\begin{figure*}[tb]
    \centering
    \includegraphics[]{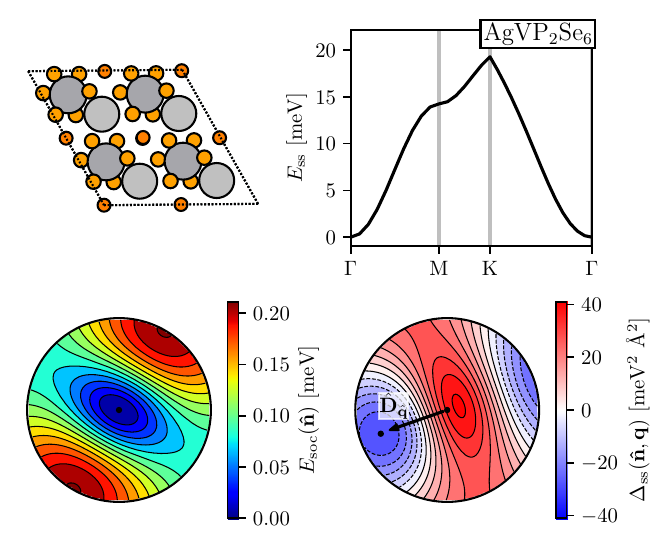}
    \caption{Analysis of the magnetic order in AgVP$_2$Se$_6$. Top left: the atomic structure, which may be regarded as a distorted honeycomb lattice formed by the V-Ag dimers. Top right: the spin spiral dispersion along the high symmetry path. Bottom left: the magnetic anisotropy, which exhibits out-of-plane easy axis and a hard axis orthogonal to the polar axis. Bottom right: the competition between DM interaction and magnetic anisotropy is dominated by the former, resulting in several directions of the normal vector where the DM spiral has lower energy than the ferromagnetic state. The direction of the DM vector $\mathbf{\hat{D}}_\mathbf{\hat{q}}$ is indicated by the arrow and the anisotropy leads to a normal vector $\mathbf{\hat{n}}$ indicated by the black dot, which is rotated slightly away (towards the atomic plane) from the direction of the DM vector.}
    \label{fig:VAgP2Se6}
\end{figure*}
In ferromagnetic crystals without inversion symmetry, the spin-orbit energy may lead to a destabilisation of the collinear order into a chiral spin spiral. In general, there will always be DM interactions favouring such states in non-centrosymmetric ferromagnets, but the formation of a spin spiral may be associated with a loss of anisotropy energy. The chiral ground state thus only occurs if the gain in DM energy is larger than the cost in magnetic anisotropy. In order to quantify when this happens we model the magnetic interactions by the Heisenberg model
\begin{align}
    H=-\frac{1}{2}\sum_{ij}J_{ij} \mathbf{S}_i\cdot\mathbf{S}_j-\frac{1}{2}\sum_{ij}\mathbf{D}_{ij}\cdot\mathbf{S}_i\times\mathbf{S}_j-\sum_{i\alpha\beta}S_i^\alpha A^{\alpha\beta}S_i^\beta,
    \label{eq:heisenberg}
\end{align}
where $\mathbf{S}_i$ is the spin operator located at site $i$, $J_{ij}$ are the isotropic exchange interactions, $\mathbf{D}_{ij}$ are the DM interactions and $A^{\alpha\beta}$ is the single-ion anisotropy tensor with $\alpha, \beta$ denoting cartesian indices.

In a classical treatment, the spin operators are replaced by axial vectors of fixed magnitude and a flat right-handed spin spiral in an arbitrary plane can be modelled as
\begin{align}
\mathbf{S}_i^{\mathbf{\hat n}}=SR_{\mathbf{\hat n}}(\cos(\mathbf{q}\cdot\mathbf{R}_i),\sin(\mathbf{q}\cdot\mathbf{R}_i), 0),
\end{align}
where $S$ is the magnitude of the spin vector, $\mathbf{R}_i$ are the unit cell vectors and $R_{\mathbf{\hat n}}$ rotates $\mathbf{\hat{z}}$ into the normal of the spiral plane $\hat{\mathbf{n}}$. Inserting this into equation \eqref{eq:heisenberg} yields the classical energy
\begin{align}
    E_\mathbf{q}^{\mathbf{\hat n}} =&-\frac{1}{2}\sum_{i}\tilde J_{0i}\cos(\mathbf{q}\cdot\mathbf{R}_i) \nonumber\\
    & - \frac{1}{2}\sum_{i}\mathbf{\hat n}\cdot\mathbf{\tilde D}_{0i}\sin(\mathbf{q}\cdot\mathbf{R}_i)-\frac{\tilde A^{xx}_{\mathbf{\hat n}}+\tilde A^{yy}_{\mathbf{\hat n}}}{2}, 
\end{align}
where we have defined the rotated anisotropy tensor $\tilde A_{\mathbf{\hat n}}=S^2R_{\mathbf{\hat n}}^TAR_{\mathbf{\hat n}}$ and absorbed the magnitude of spin into the exchange parameters such that $\tilde J_{0i}=S^2J_{0i}$ and $ \mathbf{\tilde D}_{0i}=S^2\mathbf{D}_{0i}$. If we restrict ourselves to long wavelength spirals, we can further simplify this to
\begin{align}
    E_{\mathbf{q}}^{\mathbf{\hat n}}=\frac{1}{2}\Big[\mathcal{J}_\mathbf{\hat q}\frac{q^2}{2}-J_\mathbf{0}-\mathcal{D}^\mathbf{\hat n}_\mathbf{\hat q}q-(\tilde A^{xx}_{\mathbf{\hat n}}+\tilde A^{yy}_{\mathbf{\hat n}})\Big].
    \label{eq:spiralenergy}
\end{align}
where
\begin{align}
&\mathcal{J}_\mathbf{\hat q}=\sum_{i}\tilde J_{0i}(\mathbf{\hat q}\cdot\mathbf{R}_i)^2,\qquad
J_\mathbf{0}=\sum_{i}\tilde J_{0i},\nonumber\\
&\mathcal{D}^\mathbf{\hat n}_\mathbf{\hat q}=\mathbf{\hat n}\cdot\mathbf{D}_\mathbf{\hat q},\qquad\mathbf{D}_\mathbf{\hat q}=\sum_{i}\mathbf{\tilde D}_{0i}(\mathbf{\hat q}\cdot\mathbf{R}_i).
\end{align}
We regard the model energy $E_\mathbf{q}^{\mathbf{\hat n}}$ as an approximate representation of the spin spiral energy. The exchange constant $\mathcal{J}_\mathbf{\hat q}$ can be found as twice the curvature of $E_\mathbf{q}^{\mathbf{\hat n}}$ at $\mathbf{q=0}$ and $\mathcal{D}^\mathbf{\hat n}_\mathbf{\hat q}$ is twice the slope of  $E_\mathbf{-q}^{\mathbf{\hat n}}$ (including spin-orbit coupling) at $\mathbf{q=0}$. The single-ion anisotropy tensor is obtained by fitting the magnetic anisotropy energy 
(see figure \ref{fig:soc_examples}) to the third term of Eq. \eqref{eq:heisenberg}. 

The present analysis is limited by the fact that we only calculate the spiral energies along the high symmetry path and we only calculate the DM vector $\mathbf{D}_\mathbf{\hat q}$ for $\mathbf{\hat q}$ along the two primitive reciprocal lattice vectors. There could therefore be directions in $q$-space where the DM interactions are larger, which may lead to false negatives when determining whether or not the chiral spin spirals are favoured in the analysis below. We also note that the quadratic approximation to the anisotropy limits the description to uniaxial or triaxial symmetry depending on the point group and cannot, for example, represent the in-plane anisotropy of NiBr$_2$ shown in figure \ref{fig:soc_examples}. The consequences of this approximation are further discussed in appendix \ref{app:soc-model}, where more examples of the spin-orbit energies and the resulting model energies are provided.

In order to determine the correct magnetic ground state, we compare the spiral energy \eqref{eq:spiralenergy} to the energy of a ferromagnetic ground state aligned along the axis $\hat{\mathbf{m}}$ which minimizes the single-ion anisotropy energy:
\begin{align}
    E_{\mathbf{0}}=-\frac{1}{2}\Big[J_\mathbf{0}+2{\mathbf{\hat m}}^T\cdot \tilde A\cdot\mathbf{\hat m})\Big],
\end{align}
where $\tilde A = S^2A$. In general there will exist a value of $q$ for which $E_{\mathbf{q}}^{\mathbf{\hat n}}<E_{\mathbf{0}}$ if
\begin{align}
&\Delta_{\mathrm{ss}}\equiv2\Delta A_{\mathbf{\hat n}}\mathcal{J}_\mathbf{\hat q} - (\mathcal{D}_\mathbf{\hat q}^{\mathbf{\hat n}})^2 < 0, \label{eq:condition}
\end{align}
for some pair of $\mathbf{\hat n}$ and $\mathbf{\hat q}$. Here
\begin{align}
\Delta A_{\mathbf{\hat n}}=2{\mathbf{\hat m}}^T\cdot \tilde A\cdot\mathbf{\hat m}-(\tilde A^{xx}_{\mathbf{\hat n}}+\tilde A^{yy}_{\mathbf{\hat n}}),
\end{align}
is the loss in anisotropy energy. If equation \eqref{eq:condition} is satisfied, the magnitude of the spiral ordering vector is found to be $q^{\mathbf{\hat n}}=\mathcal{D}^{\mathbf{\hat n}}_\mathbf{\hat q}/\mathcal{J}_\mathbf{\hat q}$. We may thus calculate the optimal ordering vectors (if they exist) for all directions $\mathbf{\hat n}$ and compare the resulting spiral energies from Eq. \eqref{eq:spiralenergy}, which allows us to determine the ground state. It should be noted that in some cases the conditions for forming a chiral spin spiral are trivially satisfied. For example, the case of an easy-plane magnet where any finite $z$-component of the DM vector will always favour the formation of an in-plane cycloid over the ferromagnetic state.

As an example of a predicted DM spin spiral we consider the case of AgVP$_2$Se$_6$. The parent bulk structure is a van der Waals bonded material, which was characterised in Ref. \cite{OUVRARD19881199} and is likely to be exfoliable to the monolayer limit. The structure (shown in figure \ref{fig:VAgP2Se6}) is a distorted honeycomb lattice of alternating V and Ag atoms bonded to stabilising selenium and phosphor ligands. The distortion caused by the V-Ag dimerisation introduces a polar axis and the material has been shown to comprise a three-state switchable ferroelectric corresponding to the three possible axes of dimerisation \cite{kruse2023two}. In figure \ref{fig:VAgP2Se6} we show the spin spiral dispersion along with the anisotropy and the quantity $\Delta_\mathrm{ss}$ that determines the stability of the chiral spin spiral with respect to the ferromagnetic phase. The spiral energy (without spin-orbit interaction) clearly shows a preference for ferromagnetic order and exhibits a kink at K related to the choice of band path with respect to the polar axis \cite{kruse2023two}. The magnetic anisotropy is triaxial (reflecting the two-fold symmetry), with the easy axis being out-of-plane and the hard axis being orthogonal to the polar direction. We find a sizeable DM vector, $\mathbf{D}_{\hat{\mathbf{q}}} = (6.8, 2.4, -2.5)$ meV Å, which destabilises the ferromagnet in a large region spanning roughly $\Delta \theta = 55^\circ$ by $\Delta \varphi = 88^\circ$ around the minimum at $\hat{\mathbf{n}} = (75, 200)$ as shown in figure \ref{fig:VAgP2Se6}. The triaxial magnetic anisotropy leads to an optimal direction $\mathbf{\hat n}$ that is tilted about $\Delta \theta = 5^\circ$ into the plane compared to the DM vector. The prediction of a spiral ground state in this material renders it a chiral type II multiferroic, since a change in the polar axis would induce a change in the spiral orientation.


\begin{figure}[tb]
    \centering
    \includegraphics{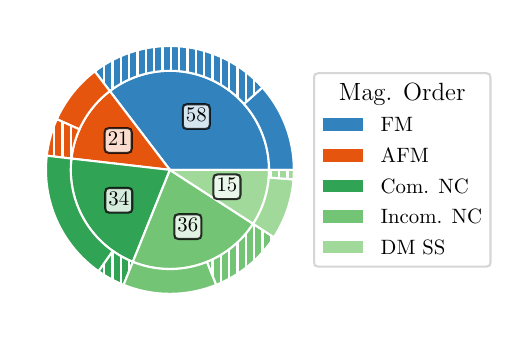}
    \caption{Distribution of magnetic order between ferromagnets (FM), collinear anti-ferromagnets (AFM), commensurate non-collinear structures (Com. NC), incommensurate spin spirals (Incom. NC) and chiral spin spirals driven by DM interactions (DM SS). The dashed area corresponds to the fraction of metals of the respective magnetic order.}
    \label{fig:distribution}
\end{figure}
\subsection{Magnetic ground states in C2DB}
The selection criteria described in section \ref{sec:workflow} yield 192 2D materials having diverse magnetic ground states with collinear as well as non-collinear order. The distribution of magnetic ground states are shown in figure \ref{fig:distribution}. We find that roughly 44\% of spin spiral calculations yield a ferromagnetic ground state, but a fifth of these get destabilised by DM interactions into chiral spin spiral ground states according to the analysis of section \ref{sec:dmi}. These materials are labeled DM spin spirals (DM SS) in figure \ref{fig:distribution} and will be discussed below. Non-collinear order, 
which encompasses both commensurate order such as $120^{\circ}$ states on triangular lattices and spin spiral order with any order parameter $\mathbf{Q}$, is found in 43\% of the 2D materials and thus occurs almost as frequently as ferromagnetic order. The remaining 13\% are collinear anti-ferromagnets which are either stripe ordered ($\mathbf{Q}=(0, 1/2)$ or $\mathbf{Q}=(1/2, 0)$) or Néel ordered ($\mathbf{Q}=(1/2, 1/2)$.

We find 15 DM spin spirals, which would all be predicted as ferromagnets in the absence of SOC. These are summarised in table \ref{tab:dmss}, where we state the spiral wavelength, spin plane orientation as well as the stoichiometry and space group (of the ferromagnetic minimal unit cell). The model used to establish the spiral ground state in these materials assumes a long wavelength spiral and the estimated wavelengths become unreliable for the materials where the wavelength is comparable to the lattice constant. These are denoted by a * in table \ref{tab:dmss}. Specifically, the linear and quadratic $q$-dependencies for DM and exchange energies in Eq. \eqref{eq:spiralenergy} respectively, are not valid if the spin stiffness $\mathcal{J}_\mathbf{\hat q}$ is sufficiently small or if the DM vector $\mathcal{D}_\mathbf{\hat q}^\mathbf{\hat n}$ is large. We show an example of this for the case of VClIO in appendix \ref{app:long_wavelength}, where we compare the linear approximation with a calculation of the DM energy along the full high symmetry path. Nevertheless, even if we are not able to provide an accurate prediction for $\mathbf{Q}$ in these materials, they will still exhibit chiral spiral order driven by DM interactions - albeit with a longer wavelength $\lambda$ than predicted here. We finally note that some of these may prefer skyrmion lattice ground states, which cannot be captured directly by spin spiral calculations and require inclusion of higher order spin interactions in a Heisenberg model description \cite{Hoffmann2020}.
\begin{table}[tb]
\begin{tabular}{lrrrr}
\hline
Formula        & $\lambda$[\AA] & $(\theta, \varphi)$ & SG           & ID      \\ \hline
CoBr$_2$       & 42             & (90, 90)            & P$\Bar{4}$m2 &         \\
NiBr$_2$       & 21             & (90, 90)            & P$\Bar{4}$m2 &         \\
TaO$_2$        & 6*             & (90, 90)            & P$\Bar{4}$m2 &         \\
CrSeI          & 138            & (90, 118)           & P3m1         &         \\
PdHfCl$_6$     & 13*            & (90, 90)            & P312         &         \\
NiHfBr$_6$     & 75             & (90, 90)            & P312         &         \\
NiHfI$_6$      & 14*            & (90, 90)            & P312         &         \\
NiZrBr$_6$     & 73             & (90, 30)            & P312         &         \\
NiZrI$_6$      & 31             & (90, 90)            & P312         &         \\
VF$_2$O        & 195            & (0, 0)              & Pmm2         &         \\
VCuP$_2$Se$_6$ & 62             & (50, 183)           & C2           &         \\
VAgP$_2$Se$_6$ & 61             & (75, 200)           & C2           & 1509506 \\
VAuP$_2$Se$_6$ & 48             & (66, 214)           & C2           &         \\
VClIO          & 4*             & (81, 100)           & Pm           &         \\
ReAu$_2$F$_6$  & 9*             & (79, 336)           & P1           &         \\
\hline
\end{tabular}
\caption{Chiral spin spiral ground states with wavelength $\lambda=q/2\pi$ and spiral plane normal axis $\hat{\mathbf{n}} = (\theta, \varphi)$. The materials marked with a * have $\lambda < 4|a|$ where $a$ is the unit cell length. This breaks the long wavelength approximation making the estimated wavelength inaccurate. We also state the space groups of the minimal non-magnetic unit cell and the COD identifier \cite{Graulis2011} of the parent bulk structure.}
\label{tab:dmss}
\end{table}

\begin{figure}[t!]
    \centering
    \includegraphics[width=\linewidth]{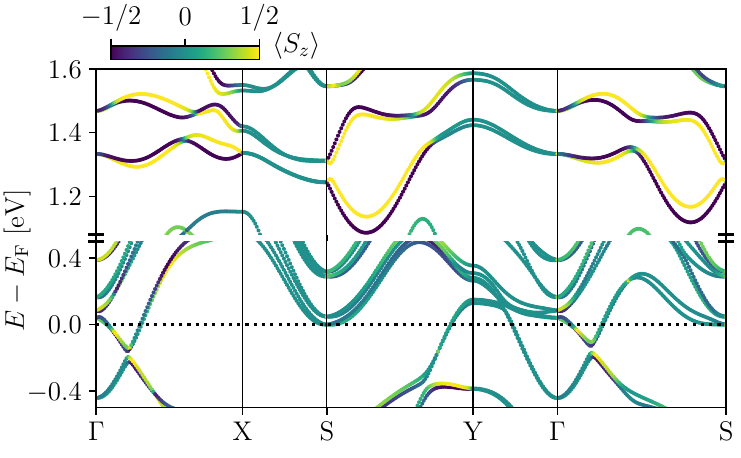}
    \caption{Band structure of MnIn$_2$Se$_4$ represented in a rectangular super cell consistent with the collinear anti-ferromagnetic order.}
    \label{fig:AFM-bs}
\end{figure}

There has recently been renewed interest in collinear anti-ferromagnets due to the fact that these may exhibit non-degenerate bands if the combination of time reversal and inversion symmetry (PT) is absent \cite{Yuan2020,Smejkal2022}. The most interesting compounds are those belonging to magnetic space groups of type III, where the band splitting originates from exchange and may become rather large. In the present work we have only considered materials with a magnetic Bravais lattice implying that any collinear anti-ferromagnet will be of type IV. Nevertheless, if PT symmetry is absent the spin-orbit interactions may still lift the band degeneracy and the band splitting may become large for materials containing heavy elements. In table \ref{tab:afm} we list the 21 materials predicted to have a collinear anti-ferromagnetic ground state and state whether or not they are invariant under PT symmetry. The majority (17) lacks PT symmetry and are expected to exhibit spin-orbit split bands. Three of these are experimentally known as bulk van der Waals bonded materials (implied by a COD/ICSD \cite{Graulis2011,Allmann2007} identifier in the table) and could be exfoliable from the parent bulk compound. In figure \ref{fig:AFM-bs} we show the the band structure of MnIn$_2$Se$_4$ obtained from a rectangular super cell containing two magnetic atoms in the predicted antiferromagnetic configuration. This compound has an out-of-plane easy axis (along $z$) and the colours signify the expectation value of $S_z$ for the individual bands. We observe a rather strong band splitting - around 50 meV in the conduction band at 1.2 eV above the Fermi level and a much smaller splitting in the vicinity of the Fermi level. The residual symmetries of the magnetic monolayer are a mirror plane orthogonal to the $x$-direction combined with either a fractional translation or time-reversal symmetry and a fractional translation combined with time-reversal symmetry. It is interesting to note that the mirror combined with translation enforces $\langle S_x\rangle=\pm\hbar/2$ for all non-degenerate bands along the invariant lines $\Gamma$-Y and X-S and these states thus have $\langle S_z\rangle=0$ despite the overall collinear magnetisation along $z$.
\begin{table}[]\label{tab:afm}
\begin{tabular}{lcrrr}
\hline
Formula          & $\mathbf{Q}$& SG           & PT    & ID     \\
\hline
MnS$_2$          & [1/2, 1/2]  & P$\Bar{4}$m2 & False &        \\
MnCl$_2$         & [1/2, 1/2]  & P$\Bar{4}$m2 & False &        \\
MnBr$_2$         & [1/2, 1/2]  & P$\Bar{4}$m2 & False &        \\
MnI$_2$          & [1/2, 1/2]  & P$\Bar{4}$m2 & False &        \\
FeCl$_2$         & [1/2, 1/2]  & P$\Bar{4}$m2 & False &        \\
FeBr2$_2$        & [1/2, 1/2]  & P$\Bar{4}$m2 & False &        \\
FeI$_2$          & [1/2, 1/2]  & P$\Bar{4}$m2 & False &        \\
CoI$_2$          & [1/2, 0.0]  & P$\Bar{3}$m1 & True  &        \\
MnAl$_2$Te$_4$   & [1/2, 0.0]  & P$\Bar{3}$m1 & True  &        \\
MnGa$_2$Se$_4$   & [1/2, 0.0]  & P$\Bar{3}$m1 & True  &        \\
MnIn$_2$Se$_4$   & [1/2, 0.0]  & P3m1         & False & 639980 \\
VBr$_2$O         & [1/2, 1/2]  & Pmm2         & False & 24381  \\
VCl$_2$O         & [1/2, 1/2]  & Pmm2         & False & 24380  \\
MnAlS$_2$I$_2$   & [1/2, 0.0]  & Pmm2         & False &        \\
CuLi$_2$O$_2$    & [0.0, 1/2]  & C2/m         & True  & 174134 \\
CrAgAs$_2$Te$_6$ & [1/2, 1/2]  & C2           & False &        \\
MnInBr$_6$       & [1/2, 1/2]  & C2           & False &        \\
MnGaO$_2$Cl$_2$  & [1/2, 0.0]  & Pm           & False &        \\
AgGeF$_6$        & [1/2, 1/2]  & P$\Bar{1}$   & False &        \\
MnGaCl$_2$Br$_4$ & [1/2, 1/2]  & P1           & False &        \\
MnGaClBr$_5$     & [1/2, 1/2]  & P1           & False &        \\
\hline
\end{tabular}
\caption{List of stable collinear anti-ferromagnets having one magnetic atom in the unit cell. In addition to the ordering vector we also state the space group of the minimal ferromagnetic unit cell (SG), whether or not the magnetic ground state has PT symmetry and the COD/ICSD identifier if a parent bulk van der Waals bonded material has been characterised experimentally.}
\end{table}

The starting point of all calculations were magnetic structures from the C2DB (obtained with PBE), which are all represented with ferromagnetic order in the minimal unit cell. In the C2DB the stability of the ferromagnetic order is roughly implied by an effective exchange coupling $J_\mathrm{nn}^\mathrm{PBE}$ obtained from the energy difference between ferromagnetic and anti-ferromagnetic configurations in $2\times1$ super cell calculations. From the present calculations we may validate how accurate this estimate is and in table \ref{tab:J_predict} we compare the PBE predictions (based on super cells) for the 67 commensurate spin spirals that had a PBE exchange constant calculated in the C2DB. In 59 cases we find agreement based on the magnetic ground state found here and $J_\mathrm{nn}^\mathrm{PBE}$ thus seems to provide a reasonable account of the stability of the ferromagnetic state present in the C2DB. For the materials predicted as commensurate spin spirals by LDA, 
we find 7 (out of 31) where the super cell calculation predict a ferromagnetic ground state with $J_\mathrm{nn}^\mathrm{PBE}>0$. In addition, we find one case (CuLi$_2$O$_2$) where PBE yield $J_\mathrm{nn}^\mathrm{PBE}<0$ (implying an antiferromagnetic state) although we obtain $q=0$ from LDA spin spiral calculations. Since $J_{nn}^\mathrm{PBE}<0$ indicates the existence of a state with lower energy than the ferromagnetic one, this can only happen as a results of different predictions for magnetic order in LDA and PBE. We note that this material is metallic and the validity of describing the magnetic interactions in terms of an effective exchange constant is questionable for metals, since the electronic structure may undergo significant changes when considering different magnetic configurations.

\begin{table}[b]
\begin{tabular}{c|c|c}
                & $J^\mathrm{PBE}_\mathrm{nn} > 0$ & $J^\mathrm{PBE}_\mathrm{nn} < 0$ \\ \hline
$q^\mathrm{LDA} = 0$   & 35   & 1     \\
$q^\mathrm{LDA} = \mathrm{HSP}$ & 7   & 24    
\end{tabular}
\caption{Comparison between predictions of the PBE nearest neighbour Heisenberg exchange 
$J^\mathrm{PBE}_\mathrm{nn}$ and LDA spin spiral ground states.}
\label{tab:J_predict}
\end{table}

\begin{figure}[t!]
    \centering
    \includegraphics[width=\linewidth]{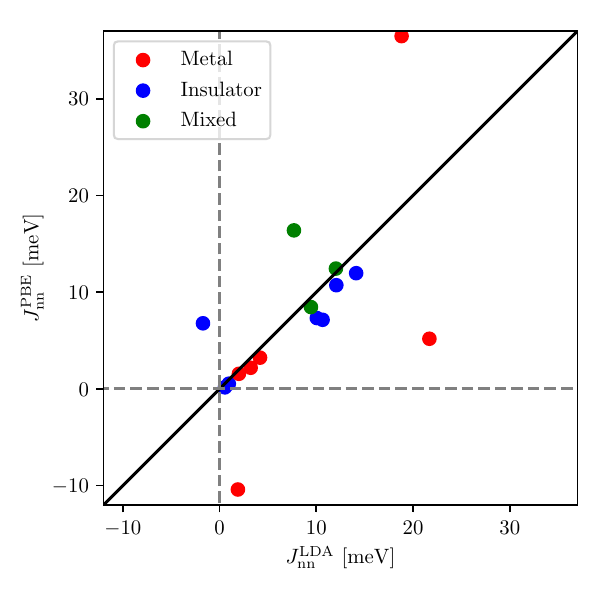}
    \caption{Comparison of LDA and PBE Heisenberg exchange for materials with an incommensurate spin spiral ground state. The calculations have been divided into metals, insulators and mixed systems where LDA and PBE disagrees on the presence of a gap in the ferromagnetic state.}
    \label{fig:exchange_predict}
\end{figure}

There is no \textit{a priori} reason to believe that PBE provides a more accurate description of magnetic order compared to LDA. But it is of course worrying if LDA and PBE can yield different predictions for the magnetic ground state. In order to compare the predictions of these two functionals we have calculated the effective exchange constants (based on a super cell approach) using LDA for the incommensurate spin spirals found in the present study. In figure \ref{fig:exchange_predict} we compare the results of LDA and PBE (taken from the C2DB) and except for a few metals we find good agreement of the sign as well as magnitude. In addition, the far majority of the exchange constants are positive, which implies that a single effective exchange interaction cannot be used as a reliable descriptor for stability in these systems. 
\begin{figure}
    \centering
    \includegraphics[width=0.4\textwidth]{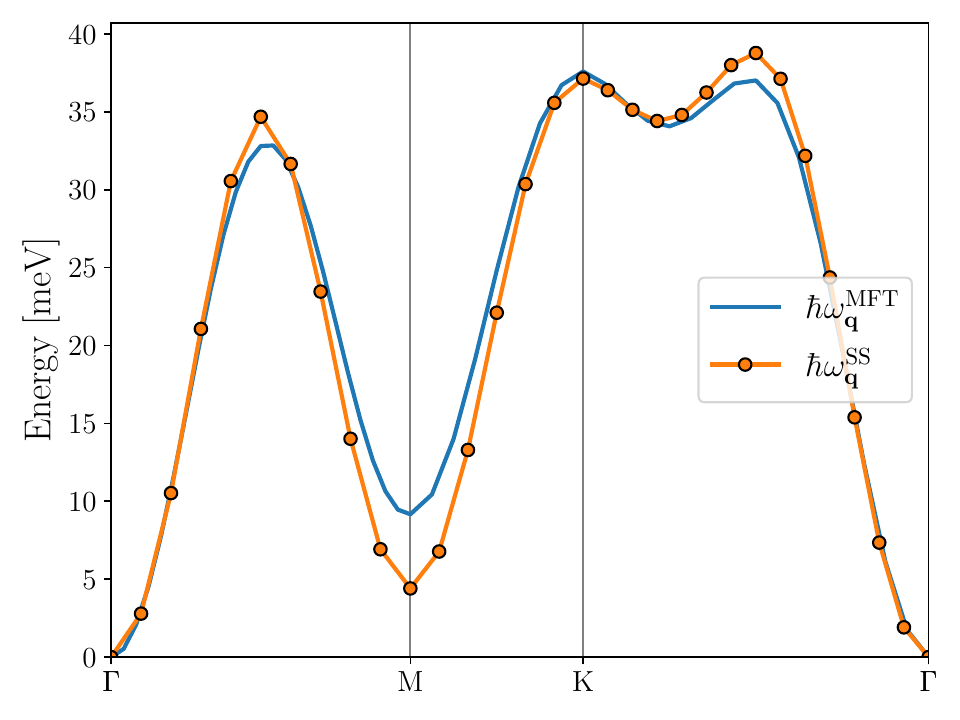}
    \includegraphics[width=0.4\textwidth]{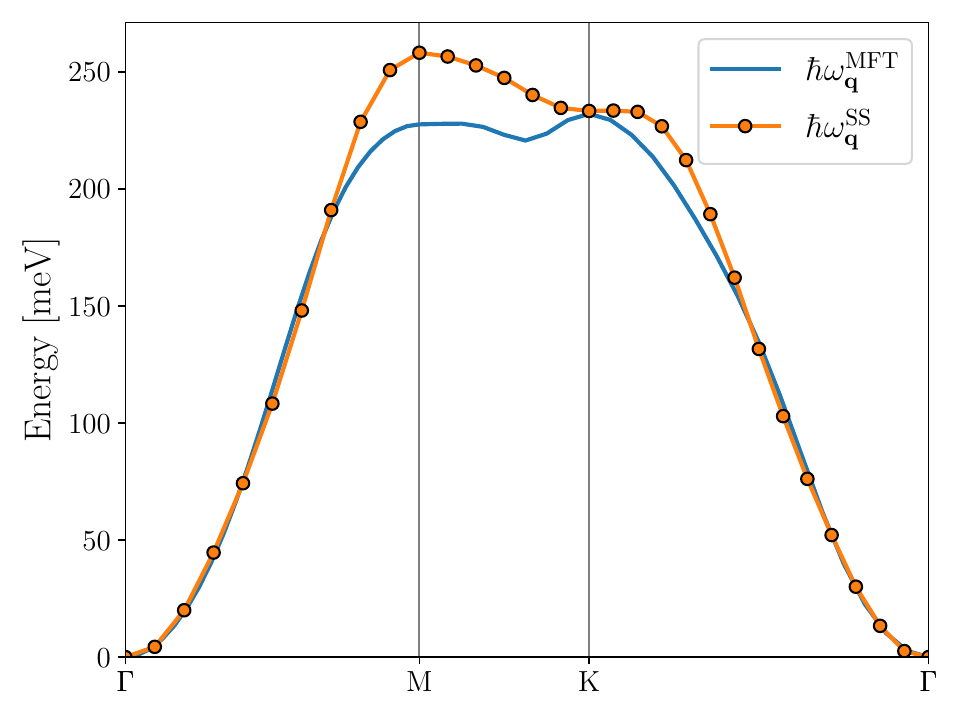}
    \caption{Comparison of the magnon dispersion in CrBrI and FeTe$_2$ obtained from spin spiral energies  ($\omega^\mathrm{SS}_\mathbf{q}$) using Eq. \eqref{eq:magnon} and energy mapping from the magnetic force theorem ($\omega^\mathrm{MFT}_\mathbf{q}$).}
    \label{fig:magnon}
\end{figure}

In general spin spiral ground states arise as a consequence of frustration between local interactions between magnetic moments. Such frustrations can occur due to the geometry of the network of interacting local moments. The case of nearest neighbour anti-ferromagnetic interactions on a triangular grid is an example of this and all the commensurate non-collinear structures in figure \ref{fig:distribution} originate from that mechanism. Alternatively, exchange frustration can occur if different interactions favour incompatible order. The simplest example of this is the spiral ground state in a linear chain of atoms that results from ferromagnetic nearest neighbour interaction and anti-ferromagnetic next nearest neighbour interaction. The abundance of positive effective exchange interactions in figure \ref{fig:exchange_predict} indicates that simple geometric frustration is not the origin of the spiralling magnetic order but rather that multiple incompatible exchange interactions contribute in determining the magnetic ground state.

\subsection{Magnons}
While the present work has focused on the ground state properties of 2D magnets, it is also straightforward to approximate the fundamental magnetic excitations (magnons) from the spin spiral energy. In the absence of spin-orbit interactions the Heisenberg model \eqref{eq:heisenberg} provides a simple relationship between the magnons and the classical spin spiral energy $E_\mathbf{q}$. The magnon dispersion for a ferromagnet with a single magnetic atom in the unit cell can be expressed as
\begin{align}\label{eq:magnon}
\hbar\omega_\mathbf{q}^{\mathrm{Magnon}}=\frac{2}{S}\Big(E_\mathbf{q}-E_\mathbf{0}\Big)
\end{align}
where $E_\mathbf{q}$ may be obtained from GBT calculations. This identification relies on the mapping from DFT to the Heisenberg model, which is expected to work well for insulators or half-metals where the electronic structure does not change much under finite spin rotations. In figure \ref{fig:magnon} we compare the magnon dispersion of CrBrI (a half-metal) and FeTe$_2$ (a metal) obtained from the GBT with a direct calculation of the exchange constants in Eq. \eqref{eq:heisenberg} using the magnetic force theorem (MFT) \cite{Durhuus2023}. In both cases we observe rather good agreement between the two approaches. We find small deviations at the Brillouin zone boundary while the dispersion in the vicinity of $\Gamma$ agrees very well. Both approaches rely on mapping DFT total energies to the Heisenberg model, but while the MFT considers infinitesimal rotations of magnetic moments the planar spin spiral calculations involve finite rotations. The deviations between the two methods can be regarded as a measure of the validity of the classical Heisenberg model when applying finite spin rotations. However, since magnons involve an infinitesimal deviation from the ground state moment the MFT must be regarded as more accurate in this context. A more rigorous treatment of magnons using the GBT would involve constrained calculations with conical spin spirals \cite{Jacobsson2013}, but we note that the good agreement in the vicinity of $\Gamma$ is expected due to the small rotations involved in the long wavelength limit. In addition, the MFT results have a slight dependence on the definition of magnetic sites that may also yield slight deviations \cite{Durhuus2023}. We also note that both approaches (as applied here) entails a systematic error related to the lack of contributions from constraining fields \cite{Grotheer2001,Bruno2003,Jacobsson2022}, but the error becomes small when the exchange splitting is much larger than the magnons band width \cite{Bruno2003}, which is the case in both of these materials.

The magnon dispersion for a general spiral ground state characterised by an ordering vector $\mathbf{Q}$ may also be obtained directly from $E_\mathbf{q}$ \cite{Toth2015}, but the procedure is somewhat more complicated and will not be pursued here.

\begin{figure}
    \includegraphics[width=\linewidth]{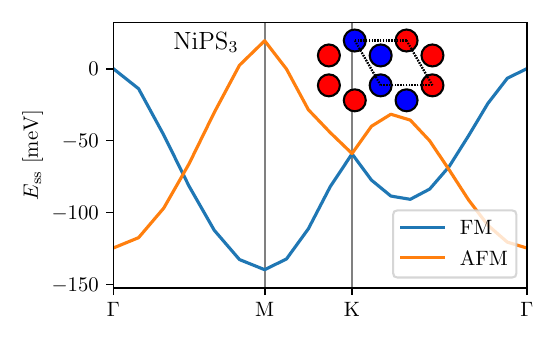}
    \includegraphics[width=\linewidth]{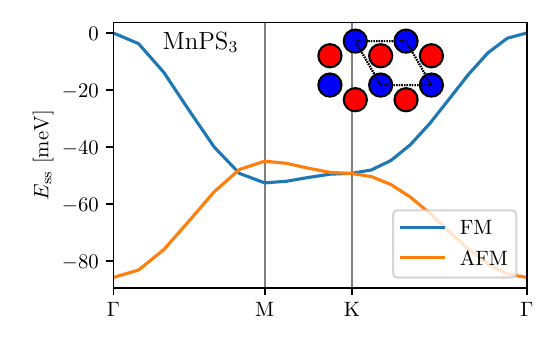}
    \caption{Spin spiral energies of the honeycomb lattices NiPS$_3$ and MnPS$_3$. The initial intracell magnetic order may be chosen as either ferromagnetic (FM) or antiferromagnetic (AFM). The two materials have ground states corresponding to the ferromagnetic M-point and the anti-ferromagnetic $\Gamma$-point respectively. These ordering vectors yield different types of collinear anti-ferromagnetic order, which are shown in the insets.}
    \label{fig:ABC3}
\end{figure}

\section{Discussion and outlook}\label{sec:discussion}


In general, the spin spiral energy landscape inherits the symmetry of the crystal and the calculated ordering vector $\mathbf{Q}$ will be degenerate with all $\mathbf{Q}'$ that are related to $\mathbf{Q}$ by symmetry. Moreover, in the absence of DM interactions $\mathbf{Q}$ and $-\mathbf{Q}$ are degenerate and may equivalently be regarded as two spirals at the same $\mathbf{Q}$ with different chiralities. In a Heisenberg model description with bilinear spin interactions, any superposition of spirals with degenerate ordering vectors have the same energy. However, higher order spin interactions \cite{Hoffmann2020} may break the degeneracy and lead to multi-$Q$ states that have lower energy. In this case the ground state may for example be a collinear \textit{uudd} anti-ferromagnet (two-$Q$ state) or have a non-coplanar skyrmion lattices (triple-$Q$ state) \cite{Kurz2001, Paul2020, gutzeit2022nano}, which are not captured by the minimal spiralling ansatz applied here. A systematic investigation of higher order interactions is beyond the scope of this work, but we note that the \textit{uudd} states have previously been evaluated for a few of the transition metal halides considered here and found to be degenerate with the corresponding single-$Q$ states. \cite{Sødequist2023}. 

The chiral spin spirals emerging from DM interaction in non-centrosymmetric ferromagnets have been investigated under the assumption of long wavelengths. However, in table \ref{tab:dmss} we find that several materials have wavelengths that are on the order of the lattice constant, which obviously contradicts the initial assumption. The calculated wavelengths are thus not reliable in these cases, but could be obtained more accurately from a comparison of the spiral energy and the DM energy along the full high symmetry path. An issue which is perhaps more critical is that the GBT is incompatible with spin-orbit interactions, which we included through the projection method. The validity of this approach has previously been established for a few particular materials 
\cite{Sandratskii2017,Sødequist2023}, but may be less accurate when the spin-orbit 
coupling is strong. A systematic comparison of projected spin-orbit with full spin-orbit (using super cells) would be highly useful in this regard.%


An additional point of concern is that the strong electronic correlation associated with $d$-orbitals may not be accurately captured by LDA and could lead to wrong predictions. In particular, we have previously shown that LDA+U yields different ordering vectors for the Mn halides, whereas other transition metal halides yield the same ordering with and without Hubbard corrections \cite{Sødequist2023}. It would be straightforward to redo the workflow with LDA+U and we anticipate to carry out such a study in the near future. Although, it is not \textit{a priori} known how accurate such calculations will be, a comparison with the bare LDA results may be used to assert whether or not the predicted magnetic order is sensitive to the inclusion of a Hubbard U.

We have only considered materials with one magnetic atom in the primitive unit cell, but the approach is in principle straightforward to generalise to materials with more magnetic atoms. This does, however, entail a few practical complications since there are different possibilities for ordering of the intracell magnetic moments at a given $\mathbf{q}$. In principle, the intracell ordering should be obtained by relaxing the electronic degrees of freedom, but in practise the convergence of the angles between individual moments may be painstakingly slow and one thus have to sample various possibilities for intracell order. Nevertheless, since the majority of the experimental work on 2D magnetism are concerned with magnetic honeycomb lattices, it seems pertinent to generalise the workflow to include such cases. For example the transition metal phosphorous tri-chalcogenides (MPS$_3$, M=Ni,Fe,Mn) have all been exfoliated to the monolayer limit and exhibits various types of anti-ferromagnetic order on the honeycomb lattice formed by the magnetic transition metal atoms \cite{Lee2016,Kim2019, Long2020}. At $\mathbf{q}=\mathbf{0}$ one may thus consider the two cases of ferromagnetic or anti-ferromagnetic ordering and these states may be extrapolated to finite ordering vectors by assuming a smooth intracell rotation of magnetic moments consistent with a chosen $\mathbf{q}$ \cite{Sødequist2023}. In figure \ref{fig:ABC3} we exemplify this for the cases of NiPS$_3$ and MnPS$_3$. In both cases we obtain a ferromagnetic as well as an anti-ferromagnetic branch corresponding to the initial intracell configurations. We find that NiPS$_3$ exhibits a minimum at $\mathbf{Q}=\mathrm{M}$ in the ferromagnetic branch corresponding to zigzag stripy order while MnPS$_3$ has the minimum at $\mathbf{Q}=\Gamma$ in the anti-ferromagnetic branch corresponding to Néel type order. Both of the results are in agreement with experiments and previous theoretical predictions based on super cell calculations \cite{Olsen2021,Kim2021}. We note that these calculations were performed with LDA+U (U=3 eV) to ensure a gapped ground state as well as better convergence and as such do not fit into the systematic workflow applied in the remainder of this work. They are merely presented here to illustrate the generalisation of the procedure to non-Bravais lattices and we leave a systematic treatment of these to future work.

We finally note, that all the results obtained in this work (ordering vectors, DM vectors, spiral energy dispersion and magnetic anisotropy) will be included in the C2DB \cite{Haastrup2018,Gjerding2021} and may be downloaded freely or browsed online.

\appendix
\section{Single-ion anisotropy approximation}
\label{app:soc-model}
\begin{figure*}
    \centering
    \includegraphics[width=0.95\textwidth]{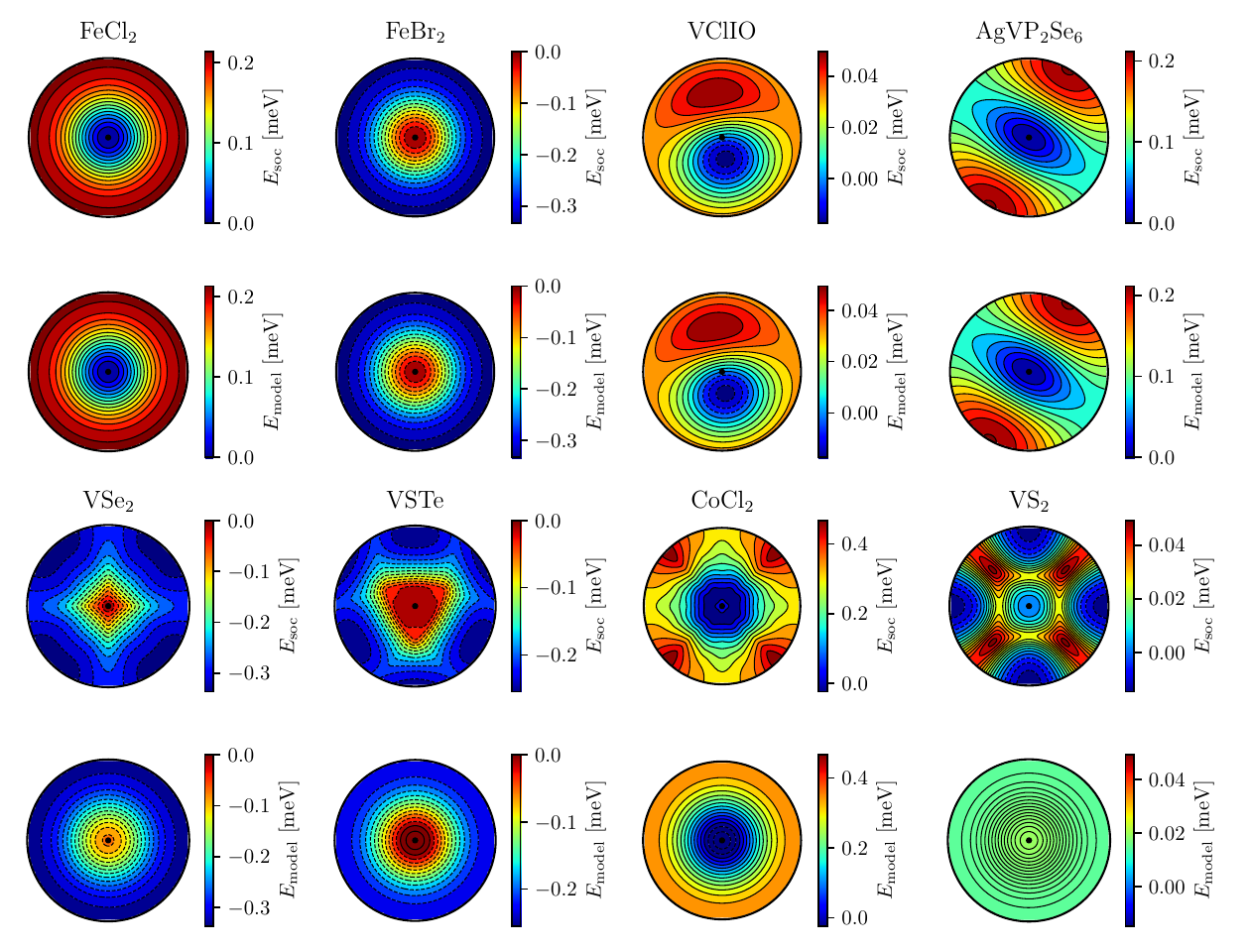}
    \caption{Examples of calculated magnetic anisotropy energies $E_\mathrm{soc}$ and the respective quadratic anisotropy model energies $E_\mathrm{model}$. Top: FeCl$_2$ (easy axis) and FeBr$_2$ (easy plane) represent uniaxial materials whereas VClIO and AgVP$_2$Se$_6$ are triaxial. These are all well reproduced by the quadratic spin model. Bottom: Examples of materials that have magnetic anisotropy with three or four-fold symmetry. These are not well described by the second order fit.}
    \label{fig:soc_model_fit}
\end{figure*}
The single-ion anisotropy tensor approximates the effects of spin-orbit coupling to second order in spin interactions on a single site. This approximation only permits uniaxial or triaxial anisotropy depending on whether the tensor has two degenerate eigenvalues or not. In particular, any $n$-fold symmetry axis with $n>2$ will enforce uniaxial symmetry and since most 2D materials that are currently scrutinised have three or six-fold symmetry axes orthogonal to the layer, these are commonly assumed to lack any magnetic anisotropy in the atomic plane. In most cases this approximation works well, but we do find a few cases where it fails rather dramatically. In figure \ref{fig:soc_model_fit} we show examples of the calculated magnetic anisotropy and the associated fit to the quadratic anisotropy model. FeCl$_2$ and FeBr$_2$ have uniaxial symmetry whereas VClIO and AgVP$_2$Se$_6$ have triaxial symmetry. In all cases the fitted anisotropy tensor yields an exact match of the magnetic anisotropy energy. The majority of materials in the present study resembles these, but we also find materials where the approximation breaks down. To exemplify such cases we show the magnetic anisotropy energy for VSe$_2$, VSeTe, CoCl$_2$ and VS$_2$, which all have three or fold-fold axes of symmetry. These show in-plane anisotropy, which are not reproduced by the quadratic anisotropy model. Such effects can only be captured by higher order anisotropic spin interactions, which might be crucial for an accurate prediction of the magnetic properties in 2D. In particular, the cases of VSe$_2$, VSeTe (and perhaps VS$_2$ depending on the approach) would be predicted to have easy planes in the quadratic model and magnetic order at finite temperature would then be excluded by the Mermin-Wagner theorem \cite{Mermin1966}. In contrast, the proper magnetic anisotropy energy exhibits in-plane easy axes, which break the continuous rotational symmetry and lead to finite Curie temperatures. 

\section{Long wavelength approximation}\label{app:long_wavelength}
\begin{figure*}[t]
    \centering
    \includegraphics[width=\textwidth]{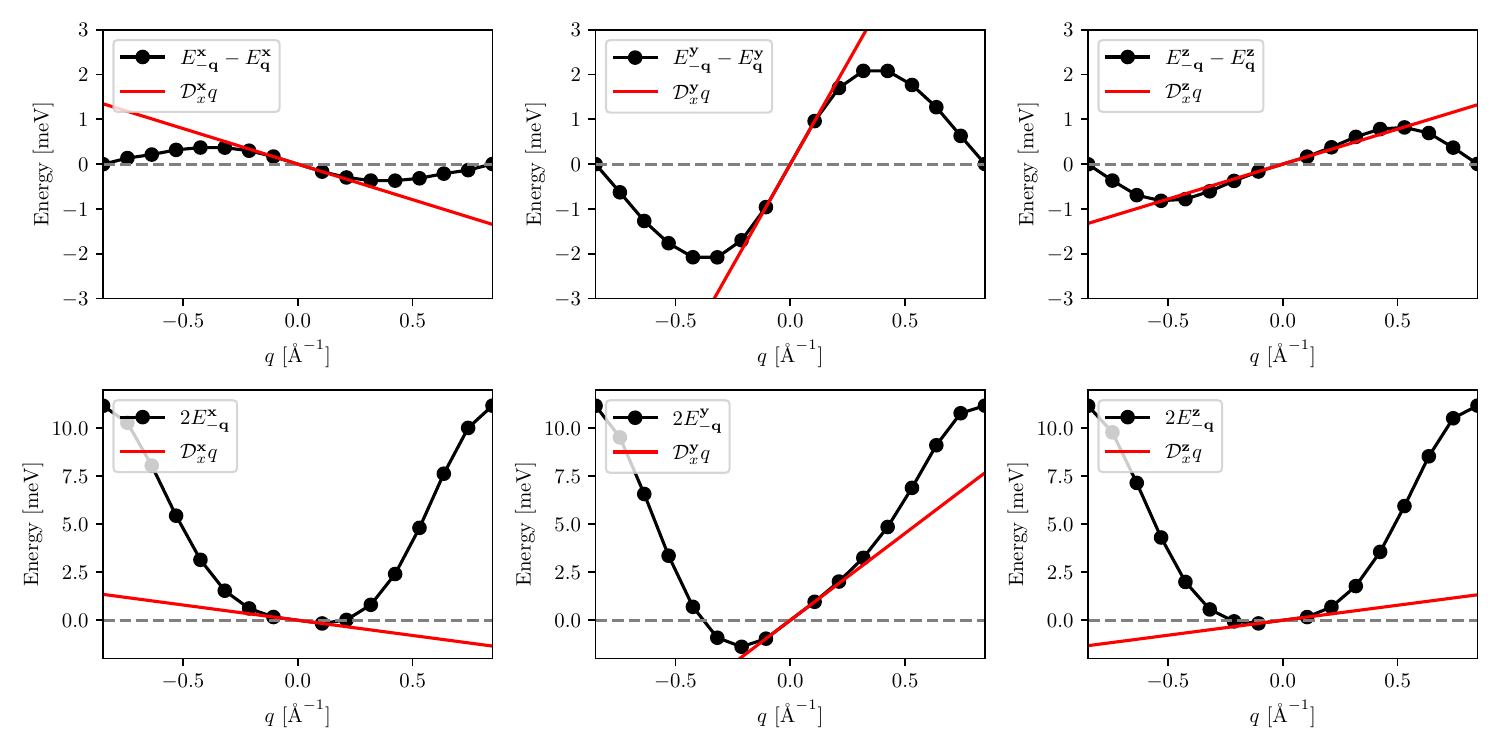}
    \caption{Top: the Dzyaloshinskii-Moriya energy of VClIO for the three spiral plane orientations along the entire path from -X to X. Bottom: the total energy resulting from the three orthogonal orientations of the spiral plane. In the absence of single-ion anisotropy, the ground state would be determined by the minimum energy.}
    \label{fig:DMI}
\end{figure*}
Several of the chiral spin spiral ground states listed in table \ref{tab:dmss} were found to have wavelengths that are on the order of the lattice vectors of the respective materials. However, the approach were based on the long wavelength limit where only the slope of the anti-symmetric spin spiral energy $\mathcal{D}$ and the spiral curvature $\mathcal{J}$ evaluated at $q=\Gamma$ entered the analysis. If the curvature is sufficiently small or if the DM vector is large this can lead to large ordering vectors that break the initial assumptions. However, one may easily generalise the approach by adding the DM interactions to the spiral energy along the entire high symmetry path. Here we exemplify this by the case of VClIO where the workflow finds an unphysical chiral order parameter outside the first Brillouin Zone due to a very small curvature at $\Gamma$. The DM vectors were determined by the workflow to be $\mathbf{D}_{\hat q_x} = (-1.6, 9.1, 1.6)$ and $\mathbf{D}_{\hat q_y} = (0.06, 0.01, 0.00)$ in [meV \AA]. In figure \ref{fig:DMI} we have calculated $E_\mathbf{q}-E_\mathbf{-q}$ between X and -X for the three different orientations of the spiral plane. 
Adding this to the spiral energy obtained without spin-orbit coupling yields the total energies shown in figure \ref{fig:DMI} (for different spiral plane orientations $\mathbf{\hat n}$) that determine the ground state order in competition with the single-ion anisotropy. Although it is certainly feasible to carry out the full analysis (including the anisotropy), the procedure is more complicated than the chosen simplified approach and we have not attempted to implement this in the workflow.


\newpage
\bibliography{bibliography.bib}

\end{document}